\documentclass[acmsmall]{acmart}

\RequirePackage{fix-cm}
\AtBeginDocument{%
  \providecommand\BibTeX{{%
    \normalfont B\kern-0.5em{\scshape i\kern-0.25em b}\kern-0.8em\TeX}}}

\setcopyright{acmcopyright}
\copyrightyear{2018}
\acmYear{2018}
\acmDOI{10.1145/1122445.1122456}

\usepackage{graphicx}

\usepackage{epstopdf}

\usepackage{framed}
\usepackage{caption}
\usepackage{subfigure} 
\usepackage{booktabs}

\usepackage[misc]{ifsym}

\usepackage{amsmath}

\usepackage{latexsym}
\usepackage{hyperref}
\acmJournal{JACM}
\acmVolume{37}
\acmNumber{4}
\acmArticle{111}
\acmMonth{8}

\begin{document}

\title{Quantum Ciphertext Dimension Reduction Scheme for Homomorphic Encrypted Data}


\author{Changqing Gong}
\affiliation{%
  \institution{Shenyang Aerospace University}
  \country{China}}
\email{gongchangqing@sau.edu.cn}

\author{Zhaoyang Dong}
\affiliation{%
  \institution{Shenyang Aerospace University}
  \country{China}}
\email{arcobelano@icloud.com}

\author{Abdullah Gani}
\affiliation{%
	\institution{University of Malaya}
	\country{Malaysia}}
\email{abdullah@um.edu.my}

\author{Han Qi}
\affiliation{%
	\institution{Shenyang Aerospace University}
	\country{China}}
\email{qihan@sau.edu.cn}

\begin{abstract}%%249 words
 At present, in the face of the huge and complex data in cloud computing, the parallel computing ability of quantum computing is particularly important. Quantum principal component analysis algorithm is used as a method of quantum state tomography. We perform feature extraction on the eigenvalue matrix of the density matrix after feature decomposition to achieve dimensionality reduction, proposed quantum principal component extraction algorithm (QPCE). Compared with the classic algorithm, this algorithm achieves an exponential speedup under certain conditions. The specific realization of the quantum circuit is given. And considering the limited computing power of the client, we propose a quantum homomorphic ciphertext dimension reduction scheme (QHEDR), the client can encrypt the quantum data and upload it to the cloud for computing. And through the quantum homomorphic encryption scheme to ensure security. After the calculation is completed, the client updates the key locally and decrypts the ciphertext result. We have implemented a quantum ciphertext dimensionality reduction scheme implemented in the quantum cloud, which does not require interaction and ensures safety. In addition, we have carried out experimental verification on the QPCE algorithm on IBM's real computing platform, and given a simple example of executing hybrid quantum circuits in the cloud to verify the correctness of our scheme. Experimental results show that the algorithm can perform ciphertext dimension reduction safely and effectively.
\end{abstract}



\maketitle

\section{Introduction}
In the information era, everyone produces a huge amount of data to be transmitted every day. In order to reduce the footprint of local memory, most users will use the cloud to save this data. In the era of cloud computing, people can operate on data in the cloud. Therefore, how to compress and extract feature data on the premise of ensuring data security in the cloud is particularly important. To ensure the safe and efficient operation of data in the cloud environment has also become the focus of many researchers. For example: Gentry proposed a breakthrough full-homomorphic encryption scheme (FHE)\cite{2009A1} has been widely studied, so that data processing rights and data ownership can be separated to prevent their own data disclosure, while making use of the computing power of cloud services. With the development of quantum computer, quantum computing will become a practical technology in the future, and quantum computing services can be provided to customers through the cloud. In order to ensure the security of quantum computing, there are mainly two kinds of quantum encryption schemes: blind quantum computing(BQC) \cite{Childs2005Secure2,2008Universal4,2013Ancilla5,2012Blind6,2012Continuous7,Vittorio2013Efficient8,Fitzsimons2017Private,2017Unconditionally} and quantum homomorphic encryption (QHE)\cite{2012Quantum9}. However, the existence of non-interactive blind quantum computing (BQC) remains to be proved.

In 2012 an important scheme is proposed by Rohde \cite{2012Quantum9}, which allows encrypted data to carry out quantum random walk. Liang Min completely defined the structure of QHE in \cite{Liang2013Symmetric10}: key generation algorithm, encryption algorithm, evaluation algorithm and decryption algorithm. Among them, the evaluation algorithm is for the operation of encrypted data in the cloud, and the key generation algorithm proposes a quantum fully homomorphic encryption (QFHE) scheme with perfect security based on quantum one-time pad,. However, the evaluation algorithm of this scheme depends too much on the key, and the server needs to receive the key transmitted by the client in order to work. Therefore, for delegated calculation, these schemes are not applicable. Tan et al proposed a QHE scheme suitable for group theory tools \cite{2014A11}, which can perform large-scale quantum computing on encrypted data. However, this scheme does not guarantee security. Because it only encrypts qubits (for encrypted data size). When  n approaches infinity, the ratio of hidden data to total data is close to 0 ($1/\log n \to 0$ ).

In 2014, Fisher et al.\cite{2014Quantum12} believe that computing power on encrypted data is a powerful tool to protect privacy. It is proved that an untrusted server can implement a set of general quantum gates on encrypted qubits without knowing any information about the input, while the client only needs to know the decryption key. The calculation results can be easily decrypted. Min Liang proposed a quantum homomorphic encryption scheme based on general quantum circuit (UQC) in \cite{Liang2015Quantum13}. In this scheme, the decryption key is different from the encryption key, and the encryption key is public. Therefore, the evaluation algorithm has nothing to do with the encryption key, so it is suitable for entrusted quantum computing between the two parties. In 2015, liang \cite{liang14} constructed a quantum version of fully homomorphic encryption and constructed two schemes. The first is the symmetric QFHE scheme, through some auxiliary qubits provided by the client, the key is a kind of quantum error correction code-CSS code. The asymmetric QFHE scheme is realized through periodic interaction between the client and the server. In 2020, a homomorphic ciphertext retrieval scheme in quantum environment is proposed \cite{2020Grover38}, and there is no need for interaction. The proposal of this scheme provides a new framework and idea for security and private protection in quantum computing.

At present, quantum machine learning is a relatively new field of development in recent years. Driven by the increasing computer ability and the progress of algorithms, machine learning technology has become a powerful tool for discovering patterns in data \cite{2011Quantum15,article16,2015An17,Wittek2014Quantum18}. The atypical patterns generated by quantum systems are considered to be inefficient in classical systems, so it is reasonable to assume that quantum computers may perform better than classical computers in machine learning tasks. The field of quantum machine learning explores how to design and implement quantum software to make machine learning faster than classical computers. Recent work has produced quantum algorithms that can be used as building blocks of machine learning programs, but the hardware and software challenges are still great. The core of machine learning algorithm and optimization algorithm is matrix calculation \cite{1989Johns19,0Convex20,book21}. The core of these algorithms is to find the eigenvalue of a matrix or the inversion of a matrix. In quantum machine learning, these steps are accelerated by using some accelerated methods of quantum information and quantum computing \cite{1996Fault22,1997A23,Chao2013Quantum24,article25,2008Quantum26,article27,2017Quantum28,2016Generic29}. For example, the Shor algorithm for solving large exponential decomposition problems \cite{1996Fault22} and the Grover algorithm for disordered database search \cite{1997A23}. Then in 2009, Harrow, Hassidim and Lloyd proposed the HHL algorithm \cite{article27}, which solves the problem of solving equations in all fields of engineering and science, and calculates on the quantum computer with as the time scale. Compared with the classical solution method, it has exponential acceleration. Many scholars have carried out real experiments on quantum computers in different ways to simply verify the algorithm and give quantum circuits \cite{2013Experimental30,2013Experimental31}. After the advent of HHL algorithm, the field of quantum machine learning has also ushered in a rich development. By using HHL as the model, a series of quantum machine learning algorithms are born.

Principal component analysis is a classical and widely used dimensionality reduction algorithm, which depends on the eigendecomposition of the covariance matrix. Therefore, quantum acceleration can be used in this field. The quantum phase estimation subroutine PhaseEstim deals with eigenvectors and eigenvalues. In 2013, Lloyd, Mohseni, Patrick proposed quantum principal component analysis (QPCA) \cite{Lloyd2014Quantum32} to prove that the density matrix of multiple copies of the system can be used to construct a unitary matrix, which can be used in data analysis, process tomography and state discrimination by using phase estimation.  and the QPCA algorithm is much faster than any known classical algorithm. But different from the classical principal component analysis algorithm, the quantum algorithm cannot achieve the dimensionality reduction function of the density matrix.

Recently, some researchers have begun to pay attention to how to implement quantum algorithms in the quantum cloud. In 2017, Huang et al.\cite{Huang2017Homomorphic} realized the homomorphic encryption experiment on the IBM quantum computing platform for the first time. This experiment homomorphically implements the quantum matrix inversion algorithm (HHL) on the IBM quantum computing platform. However, the encryption of this algorithm is not based on any quantum homomorphic encryption scheme. The security cannot be guaranteed, and the client needs to perform additional calculations on the plaintext.  In the same year, Sun et al.\cite{sun2017efficient(22)} proposed a symmetric quantum partial homomorphic encryption scheme , in which the evaluation function is independent of the key. Based on this quantum homomorphic encryption scheme, an effective symmetric searchable encryption scheme is given and proved to be secure. However, the search algorithm given in this scheme is linear search, and the efficiency will be very low when the search space is very large. 

In summary, in the future, the technology of quantum computers is truly mature, and quantum computers will not be popularized in every household. Therefore, calculations can be performed through a client-server model, and quantum homomorphic encryption will be used to perform dimensionality reduction. The calculated data is encrypted and uploaded to the quantum cloud server. After the data is processed, the server returns the ciphertext result to the client, and the client's local key is updated. There is no need to interact with the server, which can reduce the client's computing pressure.
\begin{itemize}	
	\item In this paper, this paper proposes a quantum principal component extraction algorithm (QPCE) for quantum cloud computing, The construction of the quantum circuit is completed, and a simple example is calculated on the IBMQ computing platform to verify the correctness of the proposed algorithm.
	
	\item Based on the scheme \cite{liang2020teleportation(21)}, a dimensionality reduction scheme based on quantum homomorphic encryption (QHEDR) is proposed for the first time by combining the GT-QHE scheme with our quantum dimensionality reduction algorithm in quantum cloud computing. This scheme does not require interaction to ensure security.
	
	\item It is verified that the quantum homomorphic encryption scheme is correct when there is a mixed state of multiple $T$-gates in the quantum circuit, and that the key update algorithm can satisfy the update of multiple $T$- gates.
\end{itemize}

The rest of the paper is organized as follows. We summarized some necessary preliminary knowledge of quantum computing in Sect.\ref{2}. In the section \ref{3}, we propose a quantum principal component extraction algorithm (QPCE) to extract features from the density matrix. Combined with QPCE, we propose a quantum homomorphic ciphertext dimensionality reduction (QHEDR) scheme in quantum cloud computing in Sect.\ref{4}. In the \ref{5} section, the correctness of the QPCE algorithm and the correctness of the QPCE-based quantum homomorphism dimensionality reduction (QHEDR) scheme are verified through experiments, and the experimental results are obtained. Section \ref{6} analyzes the calculation efficiency of QPCE and the security of QHEDR scheme. Finally, in Section \ref{7} we summarized the work of this article and looked forward to the work to be completed in the future.

\section{Preliminaries}
\label{2}

\subsection{Quantum computation}

The change of quantum state can be described in the language of quantum computing. Quantum computing is a new computing model that uses the laws of quantum mechanics to calculate quantum information. In quantum computing, the computational model of quantum circuit is usually composed of wires and basic quantum gates. The classical Clifford quantum bit gate is divided into single quantum bit gate and multi-quantum bit gate. The single quantum bit gate includes: pauli $X$ and $Z$, Hadamard gate $H$ and phase gate $S$. The multi-quantum bit gate include: $CNOT$ gate, $SWAP$ gate. For a detailed introduction to quantum computing gate, please refer to \cite{2011Quantum15}.
 In addition to the Clifford gate, there is only one non-Clifford gate
$T = \left( {\begin{array}{*{20}{c}}
	1&0\\
	0&{{e^{i\frac{\pi }{4}}}}
	\end{array}} \right)$.
There is at least one non-Cifford gate in every circuit, so it is very important in quantum computation. The conjugate matrix of T gate is
${T^\dag } = \left( {\begin{array}{*{20}{c}}
	1&0\\
	0&{{e^{ - i\frac{\pi }{4}}}}
	\end{array}} \right)$.
In the QHE scheme and quantum principal component extraction algorithm we use, the core is the update of T-gate and the construction of controlled U-gate.

\subsection{Quantum Teleportation}

First of all, the EPR pair is an entangled quantum state. Four Bell states can be expressed as:
\begin{equation}
\left| {{\beta _{ab}}} \right\rangle  = \left( {{Z^b}{X^a} \otimes I} \right)\left| {{\beta _{00}}} \right\rangle 
\end{equation}

\begin{figure}[htbp]
	\centering
	\subfigure[Quantum circuit to create Bell states]{
		\includegraphics[width=2.0in]{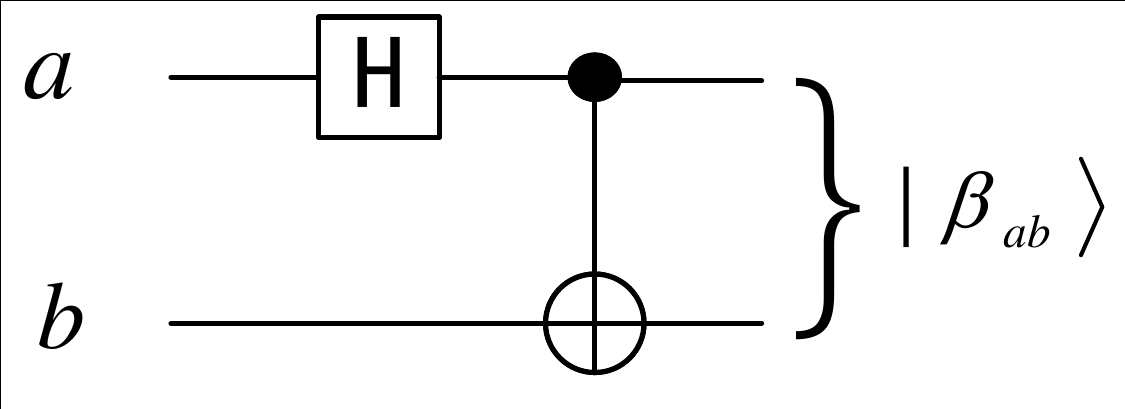}
	}
	\subfigure[Quantum circuit for of teleportation a qubit.]{
		\includegraphics[width=2.5in]{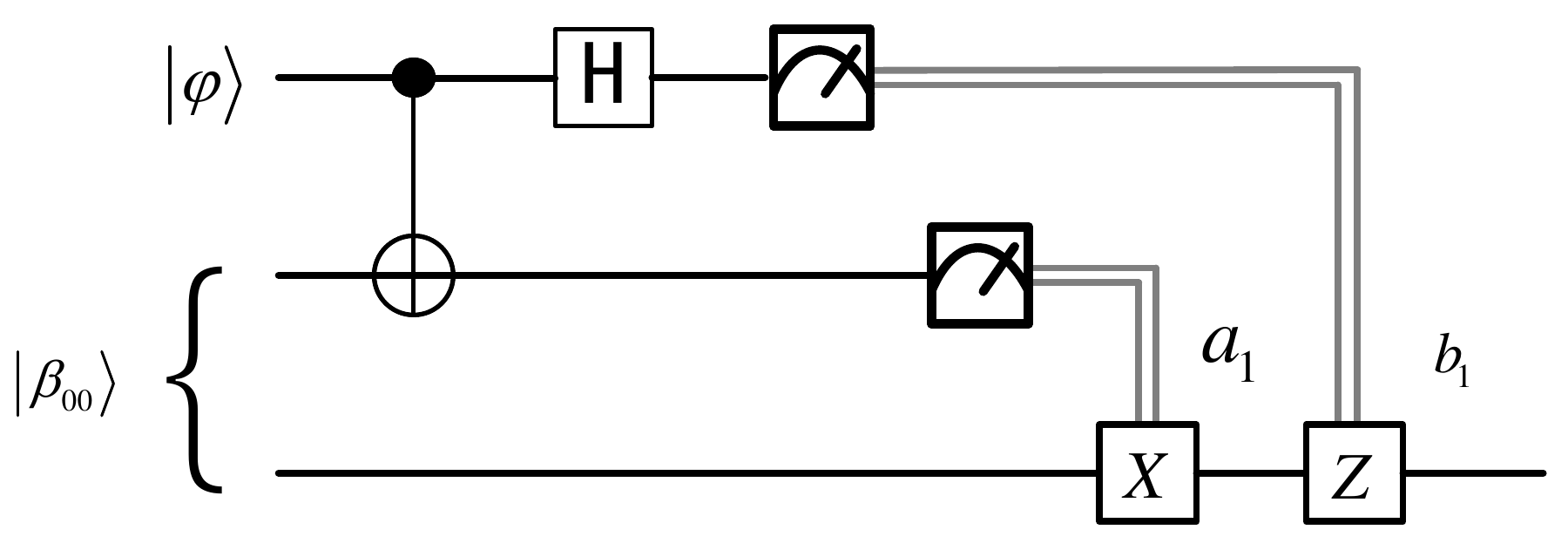}
	}
	\caption{Two examples of quantum teleportation.}
	\label{figure1}
\end{figure}

Which $a,b \in \left\{ {0,1} \right\}\,\;,\left| {{\beta _{00}}} \right\rangle  = \frac{1}{{\sqrt 2 }}\left( {\left| {00} \right\rangle  + \left| {11} \right\rangle } \right)$.
It can be generated by a Hadamard gate and a CNOT gate. The following in fig.\ref{figure1}(a) shows the quantum circuit.

Quantum teleportation is one of the most amazing applications of quantum physics in the field of information theory. Quantum teleportation is a method of moving quantum state without even a quantum communication channel connection between the sender and the receiver. Even when Alice only sends classical information to Bob, it allows quantum information to be transmitted from Alice to Bob. Its quantum circuit is shown in the following figure \ref{figure1}(b):

Through quantum teleportation, the concept of " Rotated Bell basis " is proposed \cite{Jozsa2005AnIT34}. For any single-bit U-gate, the quantum state that defines " U-rotating Bell basis " can be expressed as
\begin{equation}
\left| {\beta {{\left( U \right)}_{ab}}} \right\rangle  = \left( {{U^\dag } \otimes I} \right)\left| {{\beta _{ab}}} \right\rangle  = \left( {{U^\dag }{Z^b}{X^a} \otimes I} \right)\left| {{\beta _{00}}} \right\rangle
\end{equation}
For a single qubit, the quantum state after this operation is as follows:
\begin{equation}
\left| \varphi  \right\rangle  \otimes \left| {{\beta _{00}}} \right\rangle  = \sum\limits_{a,b \in \left\{ {0,1} \right\}} {\left| {\beta {{\left( U \right)}_{a,b}}} \right\rangle }  \otimes {X^a}{Z^b}U\left| \varphi  \right\rangle
\end{equation}
According to this definition, 'U-rotate Bell basis'. Liang \cite{2019QuIP...19...28L35} proposed a quantum homomorphic encryption scheme based on teleportation.

\subsection{Key update algorithm}
\label{2.3}
In this scheme, the client and server use Pauli X and Pauli Z gate to implement the encryption and decryption protocol for plaintext and ciphertext.
\begin{equation}
{X^a}{Z^b}\rho  \to \varphi
\end{equation}

Where $\rho $ represents the density matrix of plaintext, $\varphi $ is the encrypted density matrix $a,b$ represents the encryption key randomly selected from $\left\{ {0,1} \right\}$

\begin{figure}
	\includegraphics[scale=0.4]{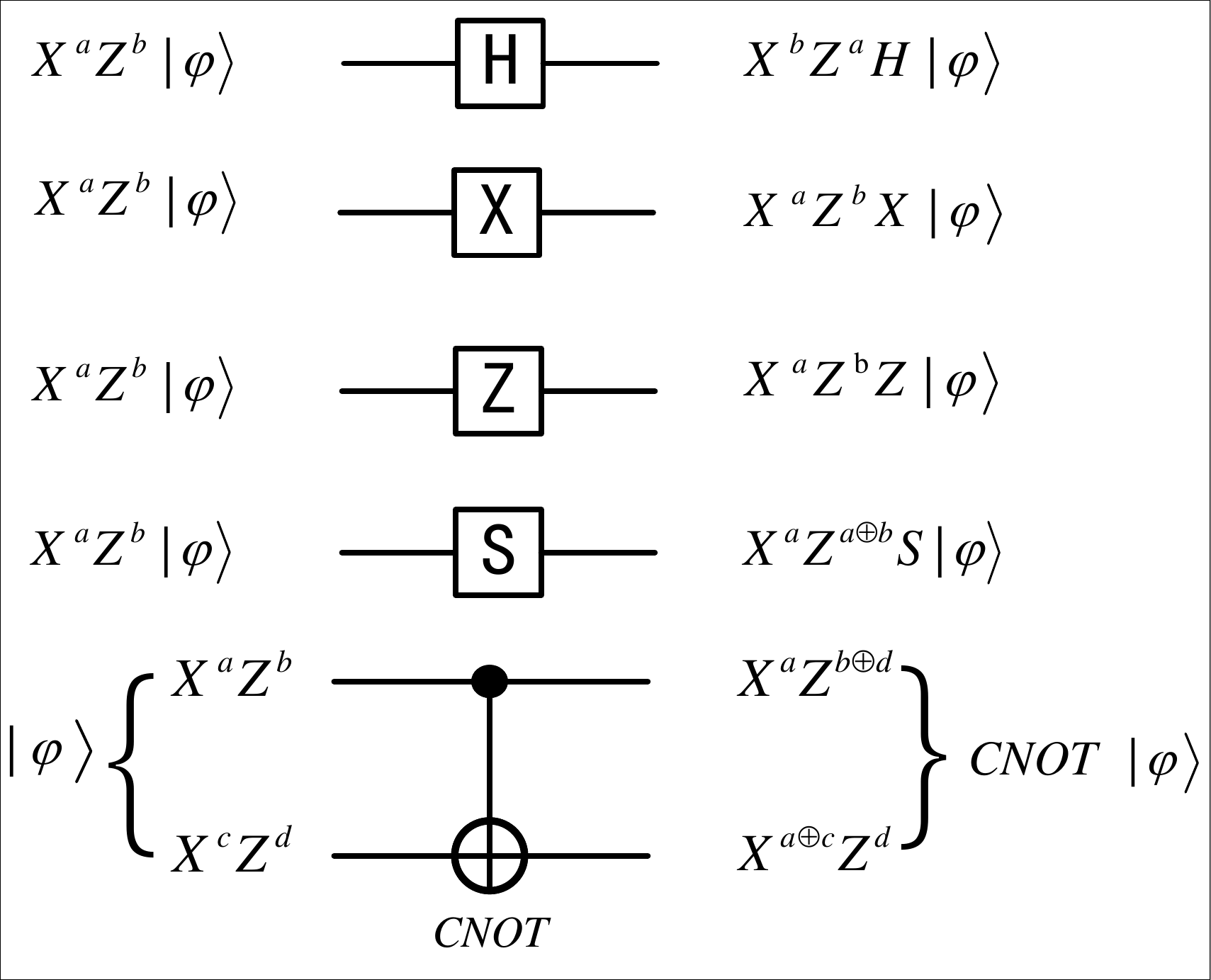}
	\caption{Key update proces of Clifford gate. Including Pauli $H$, $X$, $Z$, $S$, $CNOT$ gate key update process.}
	\label{figure3}
\end{figure}

In order to ensure the security of the encryption scheme, Boykin and Roychowdhury \cite{2003Optimal36} introduced a secure and feasible quantum one-time pad(QOTP). As long as the keys $a$ and $b$ are randomly generated to make $\left( {a,b} \right) \in \left\{ {0,1} \right\}$, and use it only once. Then the scheme is perfectly safe.Therefore, according to the exchange rules between the Clifford gate and the Pauli matrix, the secret key update rule as shown in the figure \ref{figure3} is established.

\begin{figure}[h]
	\centering
	\includegraphics[scale=0.4]{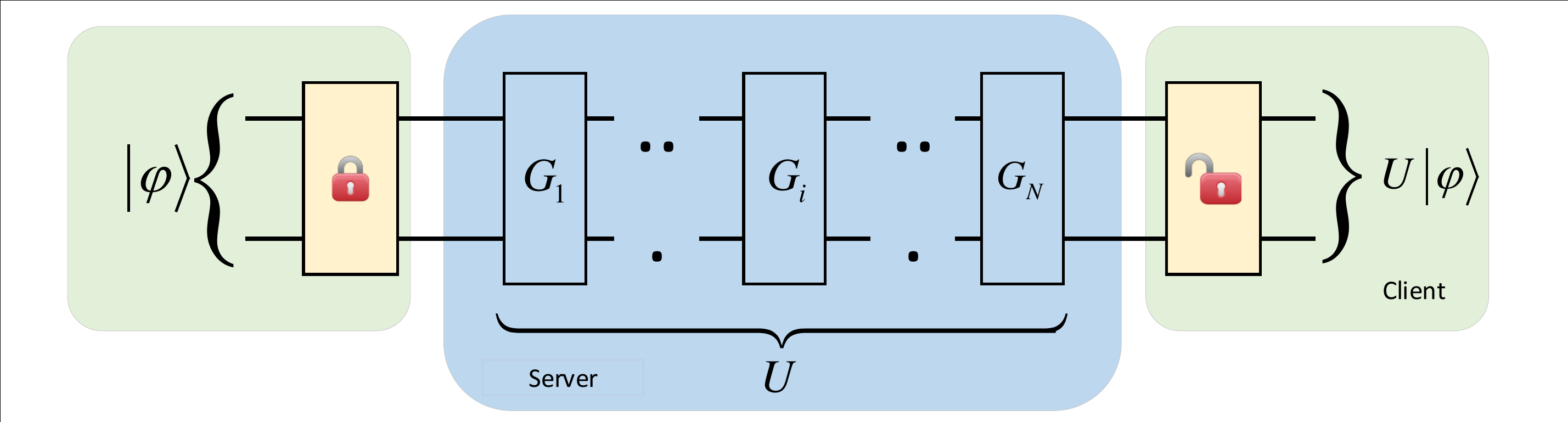}
	\caption{The composition of Quantum Computing Operation $U$.}
	\label{figure4}
\end{figure}

The encrypted quantum state $\left| \varphi  \right\rangle $ is sent by the client to the server.The server performs quantum computation $U$ on $\left| \varphi  \right\rangle $, which can be regarded as composed of various quantum gates of $G\left[ 1 \right]$,$G\left[ 2 \right]$,...,$G\left[ N \right]$, as shown in figure \ref{figure4}.

When $G \in \left\{ {X,Z,H,S,CNOT} \right\}$ , the client updates the secret key according to the operation of the server and the secret key update algorithm, and then uses the final updated secret key to decrypt the ciphertext to get $U\left| \varphi  \right\rangle $. However, when $G \in \left\{ {T,{T^\dag }} \right\}$ and key $a = 1$,an error occurs, and in order to eliminate this error, the client introduces ${\left| {{\beta _{00}}} \right\rangle _{{\rm{sc}}}}$ and measures its '${S^a}$ rotation'.The update process is shown in figure \ref{figure5}:

\begin{figure}[h]
	\includegraphics[scale=0.6]{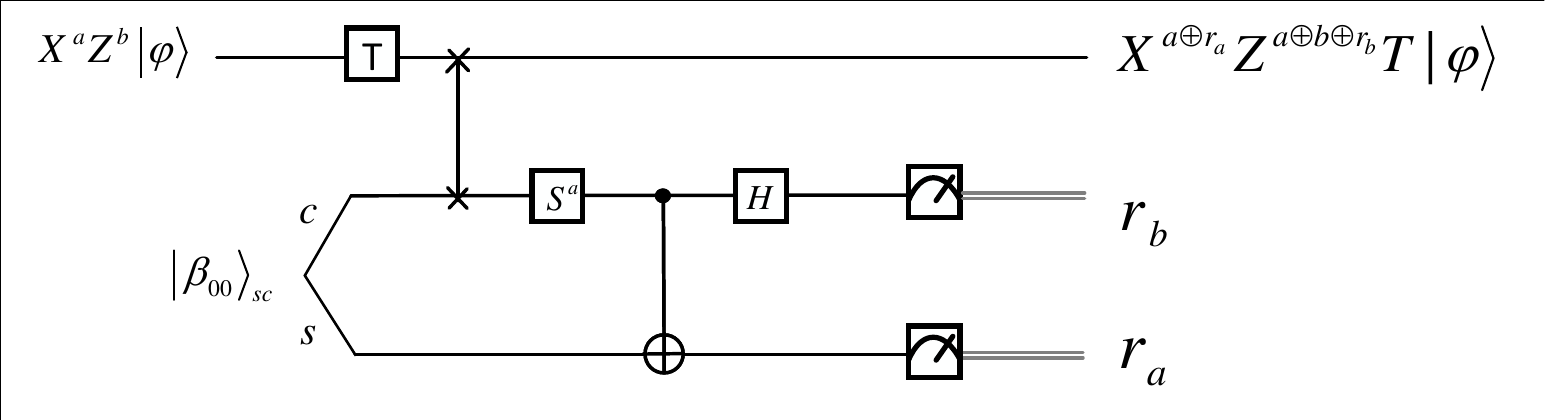}
	\caption{The process of updating the secret key of the T gate.}
	\label{figure5}
\end{figure}

The client performs ${S^a}$ rotation measurement on ${\left| {{\beta _{00}}} \right\rangle _{{\rm{sc}}}}$ to get the measurement result ${r_a},{r_b}$. And update the secret key according to the update formula. This scheme can be used to evaluate the homomorphism of general circuits. the ${S^a}$ rotation measurement can be postponed until other quantum circuits are completed, and the measurement base $\Phi \left( {{S^a}} \right)$ only depends on the secret key $a$.The specific update steps of the T gate of the update algorithm can be found in \cite{2019QuIP...19...28L35,2003Optimal36}.

\section{Quantum principal component extraction algorithm}
\label{3}
In this secrtion, we will explain the main steps of the quantum principal component extraction algorithm (QPCE) and the effect of its implementation. We give $n$ copies of a density matrix $ \rho $. The quantum phase estimation algorithm can be used to calculate the eigenvalues and Eigenvectors of the matrix. However, the quantum phase estimation algorithm requires the unitary matrix $U={e^{ - i\rho t}}$ as the input. An operation enables us to apply the construction of unitary gate ${e^{ - i\rho t}}$ of any Hermitian matrix to any density matrix $\sigma $.
\begin{equation}
\begin{array}{l}
t{r_P}{e^{ - iS\Delta t}}\rho  \otimes \sigma {e^{iS\Delta t}} = \left( {{{\cos }^2}\Delta t} \right)\sigma  + \left( {{{\sin }^2}\Delta t} \right)\rho  - i{\sin ^2}\Delta t\left[ {\rho ,\sigma } \right]\\
\quad \quad \quad \quad \quad \quad \quad {\kern 1pt}  = \sigma  - i\Delta t\left[ {\rho ,\sigma } \right] + O\left( {\Delta {t^2}} \right)
\end{array}
\end{equation}
In short, for the given $n$ copies, the unitary gate ${e^{ - i\rho t}}$ can be obtained by repeating ep.6 $n$ times. This method was proposed by Lloyd et al. in \cite{Lloyd2014Quantum32}.

We input the quantum state of the density matrix $\left| {{\varphi _\rho }} \right\rangle $ , the unitary matrix ${e^{i\rho {t_0}}}$ and a threshold constant $\tau $, and output the quantum state. 
\begin{equation}
\left| {{\varphi _s}} \right\rangle {\rm{ = }}\sum\limits_{k = 1}^s {\left( {{\lambda _k} - \tau } \right)} \left| {{e_k}} \right\rangle \left| {{v_k}} \right\rangle
\label{eq.7}
\end{equation}
Next we will describe the specific details of the algorithm.

\subsection{Quantum state preparation}

The steps for preparing the quantum state to be input are as follows:

\textbf{Step1:} To standardize and normalize the classical data, where each sample ${v_k}$ with N-dimensional vector,$k = \left\{ {1,...,M} \right\}$. First of all, we should subtract average value $\overline v $ from each training vector, and divide it by the vector norm to normalize it to construct an appropriate quantum state.
\begin{equation}
{v_k} \to {v_k} - \overline v ,{v_k} \to {\left| {{v_k}} \right|^{ - 1}}{v_k}
\end{equation}

In addition, the processed ${v_k}$ can also be standardized to obtain the correlation matrix, but this is not necessary.

\textbf{Step2:} The $i = \left\{ {1,...,N} \right\}$ components of the classical training vector $v$ are encoded into quantum states. i.e. $v \to \left| v \right\rangle  = \sum\limits_{i = 1}^n {{v_i}} \left| i \right\rangle $, where the quantum state is accurately defined and the probability addition is 1. N-dimensional vectors can be encoded into $n = {\log _2}N$ qubits.

\textbf{Step3:} The covariance matrix is represented by the density matrix, and the mixed state is represented by the density matrix $\rho {\rm{ = }}\frac{1}{M}\sum\limits_{k = 1}^M {\left| {{v_k}} \right\rangle } \left\langle {{v_k}} \right|$. The tensor product $\left| {{v_k}} \right\rangle \left\langle {{v_k}} \right|$ can be expressed as a matrix
\begin{equation}
\left| {{v_k}} \right\rangle \left\langle {{v_k}} \right|{\rm{ = }}\left[ {\begin{array}{*{20}{c}}
	{v_1^{(k)}v_2^{(k)}}&{v_1^{(k)}v_2^{(k)}}& \cdots &{v_1^{(k)}v_N^{(k)}}\\
	{v_2^{(k)}v_1^{(k)}}&{v_1^{(k)}v_1^{(k)}}& \cdots &{v_2^{(k)}v_N^{(k)}}\\
	\vdots & \vdots &{}& \vdots \\
	{v_N^{(k)}v_1^{(k)}}&{v_N^{(k)}v_2^{(k)}}& \cdots &{v_N^{(k)}v_N^{(k)}}
	\end{array}} \right]
\end{equation}

So we get the sum of the tensor product:
\begin{equation}
\frac{1}{M}\sum\limits_{k = 1}^M {\left| {{v_k}} \right\rangle \left\langle {{v_k}} \right|{\rm{ = }}\frac{1}{M}\left[ {\begin{array}{*{20}{c}}
		{\sum\nolimits_k {v_1^{(k)}v_2^{(k)}} }&{\sum\nolimits_k {v_1^{(k)}v_2^{(k)}} }& \cdots &{\sum\nolimits_k {v_1^{(k)}v_N^{(k)}} }\\
		{\sum\nolimits_k {v_2^{(k)}v_1^{(k)}} }&{\sum\nolimits_k {v_1^{(k)}v_1^{(k)}} }& \cdots &{\sum\nolimits_k {v_2^{(k)}v_N^{(k)}} }\\
		\vdots & \vdots &{}& \vdots \\
		{\sum\nolimits_k {v_N^{(k)}v_1^{(k)}} }&{\sum\nolimits_k {v_N^{(k)}v_2^{(k)}} }& \cdots &{\sum\nolimits_k {v_N^{(k)}v_N^{(k)}} }
		\end{array}} \right]}
\end{equation}

Therefore, for the reduced-dimensional data, it is equivalent to the covariance matrix, and the classical data is also standardized in the first step of the algorithm, which can be equivalent to the corresponding density matrix of the correlation matrix.

\subsection{Implementation steps of Quantum principal component extraction}
Secondly, the whole process of dQuantum principal component extraction can be expressed as follows:
\begin{equation}
{U_{QPCE}} = \left( {{I^a} \otimes U_{PE}^\dag } \right)\left( {{U_R} \otimes {I^b}} \right)\left( {{I^a} \otimes U_{PE}^{}} \right)
\end{equation}

Where ${U_{{\rm{PE}}}}$ represents phase estimation and ${U_R}$ represents controlled rotation. $U_{PE}^\dag $ stands for the inverse operation of ${U_{{\rm{PE}}}}$. Obviously, the quantum principal component extraction algorithm (QPCE) is similar to the matrix inversion algorithm (HHL algorithm). However, due to the functions of QPCE and HHL are different, ${U_{{\rm{PE}}}}$ and ${U_R}$ need to be improved.

The quantum subroutine ${U_{{\rm{PE}}}}$ can be expressed as:
\begin{equation}
\begin{array}{l}
{U_{PE}} = {U_{PE}}\left( \rho  \right)\\
\quad \,\,{\kern 1pt} {\kern 1pt} \; = \left( {F_{\rm{T}}^\dag  \otimes {I^B}} \right)\left( {{{\sum\limits_{\tau  = 0}^{T - 1} {\left| \tau  \right\rangle \left\langle \tau  \right|} }^C} \otimes {e^{i\rho \tau {t_0}/T}}} \right)\left( {{{\rm{H}}^{ \otimes t}} \otimes {I^B}} \right)
\end{array}
\end{equation}

In left-to-right order, $H$ is the precision represented by the Hadamard gate acting on register $B$ and $t$ represents the accuracy represented by phase estimation. 

The second part is a part of the controlled U gate, which indicates that the phase estimation operation is performed on the characteristic space of $\rho $\cite{article27}, The mathematical form can be expressed as eq.13. 
\begin{equation}
{\sum\limits_{\tau  = 0}^{T - 1} {\left| \tau  \right\rangle \left\langle \tau  \right|} ^C} \otimes {e^{i\rho \tau {t_0}/T}}
\end{equation}

$F_{\rm{T}}^\dag$ is the inverse transform of quantum fourier transform \cite{2011Quantum15}. At this time, through the phase estimation algorithm, the current state is changed to:
\begin{equation}
{\rm{tr}}\left( {\left| {{\varphi _1}} \right\rangle \left\langle {{\varphi _1}} \right|} \right){\rm{ = 1/}}{{\rm{m}}^2}\sum\limits_{k = 0}^m {{v_k}^2\left| {{\lambda _k}} \right\rangle \left\langle {{\lambda _k}} \right| \otimes \left| {{e_k}} \right\rangle \left\langle {{e_k}} \right|}
\end{equation}

${U_R}$ indicates that the purpose of controlled rotation is to redistribute the proportion 
\begin{equation}
\left| {{\varphi _1}} \right\rangle  = 1/m\sum\limits_{k = 0}^m {{\lambda _k}} \left| {{e_k}} \right\rangle \left| {{v_k}} \right\rangle
\end{equation}
of each feature in the quantum state, and the method of dimensionality reduction by threshold is also realized here. It can be expressed as:
\begin{equation}
\left| 0 \right\rangle \left| z \right\rangle  = \left( {\frac{{\eta \left( {\sqrt z  - \tau } \right)}}{{\sqrt z }}\left| 1 \right\rangle  + \sqrt {1 - \frac{{{\eta ^2}{{\left( {\sqrt z  - \tau } \right)}^2}}}{z}} \left| 0 \right\rangle } \right)\left| z \right\rangle
\end{equation}

where $\sqrt z  > \tau $ . By changing the probabilistic amplitude of each ground state $\left| {{e_k}} \right\rangle \left| {{v_k}} \right\rangle $ from ${\lambda _k}$ to ${\left( {{\lambda _k} - \tau } \right)_ + }$, it is realized by a transformation ${\left( {{\lambda _k} - \tau } \right)_ + } = {\lambda _k} \times \frac{{{{\left( {\sqrt {\lambda _k^2}  - \tau } \right)}_ + }}}{{\sqrt {\lambda _k^2} }}$. The specific other implementation steps are similar to the implementation of the QSVT algorithm \cite{2018Efficient37}.

Finally, we get the density matrix $\left| {{\varphi _{\rm{s}}}} \right\rangle $ in eq.\ref{eq.7} after dimensionality reduction. At this time, we can decoherent the system through inverse quantum phase estimation, and what is obtained after decoherence is the reduced-dimensional density matrix.

\section{A dimensionality reduction scheme base on QHE in quantum cloud computing}
\label{4}
In this part, we will propose the details of our dimensionality reduction scheme based on QHE (QHEDR) . We use the quantum principal component extraction (QPCE) proposed in the Sect.\ref{3} to realize the dimensionality reduction function in the ciphertext environment in quantum cloud computing. In our scheme, we assume that Alice is the client and bob is the server. If Alice's computing power is limited, Bob is required to do a matrix dimensionality reduction in the cloud. And you need to make sure that Bob doesn't know the specific solution. Quantum principal component analysis algorithm and quantum homomorphic encryption algorithm can ensure the speed and security. Therefore, we combine quantum principal component extraction algorithm with quantum homomorphic encryption to produce a new quantum dimensionality reduction scheme for ciphertext data. Alice overlays and encrypts plaintext and sends the ciphertext to Bob. Then Bob carries out the dimensionality reduction operation of principal component extraction on the ciphertext in the cloud. And the measurement results are returned to Alice. Alice updates the key according to the process of quantum principal component analysis and decrypts the measurement results. That is, the final result of dimensionality reduction can be obtained.

Quantum circuit $C$ can be composed of six parts:\textbf{preparation state, secret key generation, encryption, evaluation function, measurement and decryption}. The general process of the scheme is shown in the figure \ref{figure6}:

\begin{figure*}
	\includegraphics[scale=0.7]{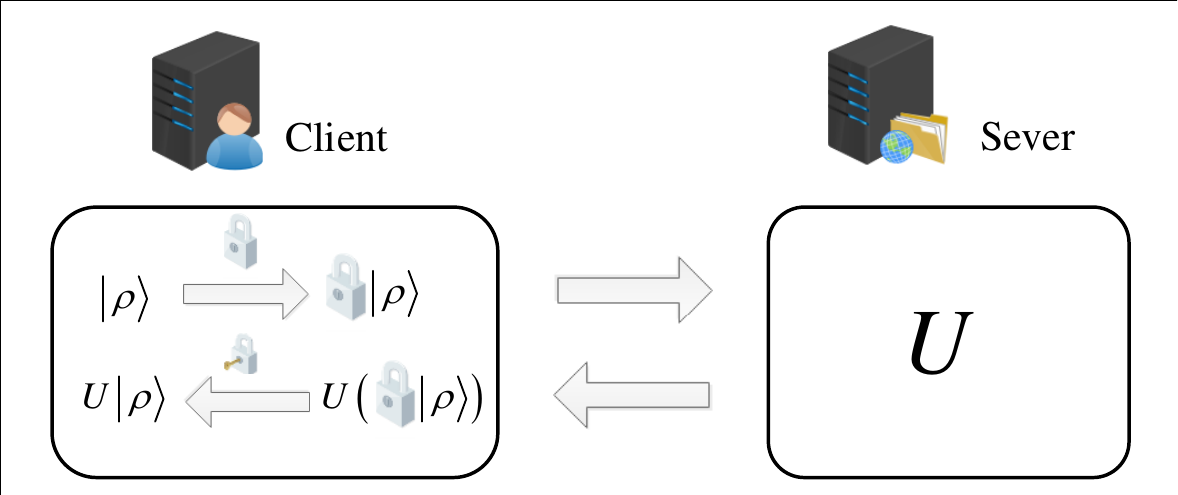}
	\caption{The specific solution of Quantum homomorphic encryption Scheme. The interaction process does not need to upload the key.}
	\label{figure6}
\end{figure*}

Next, we will introduce the detailed scheme:

	\textbf{1. State preparation}: Alice standardizes a set of N-dimensional data ${v_k}$ to obtain training vectors, in which $k \in \left\{ {1,...,N} \right\}$. The density matrix P is formed and quantized. And the $M$ group Bell state $\left\{ {{{\left| {{\beta _{00}}} \right\rangle }_{{c_i},{s_i}}},i = 1,...,M} \right\}$ is generated as an auxiliary quantum state to be saved by the client and the server respectively.
	
	\textbf{2. Generate secret key}:Alice randomly generates a pair of 2n bits of initial key $ek = \left( {{a_0},{b_0}} \right) \in {\left\{ {0,1} \right\}^n}$
	
	\textbf{3. Encryption}: $Enc\left( {ek,\rho } \right)$. For $n$-bit density matrix clients that need to encrypt, use the secret key to perform one secret operation at a time.$\varphi  \to Enc\left( {ek,\rho } \right){\rm{ = }}{X^{{a_0}}}{Z^{{b_0}}}\rho {Z^{{b_0}}}{X^{{a_0}}}$ And send the ciphertext result $\varphi $ to Bob.
	
	\textbf{4. Evaluation function}: $Eval\left( {C,\left\{ {{s_i},{c_i}} \right\}_{i = 1}^M,\varphi } \right)$. The server needs to undertake two tasks here.
	
	\textbf{4.1. Quantum principal component extraction }: With the help of the auxiliary qubit ${s_i}$, the server performs the operation of the quantum circuit $C$ of the Quantum principal component extraction. The QPCE performs N-step gate operation on the encryption matrix $\varphi $ as $G\left[ 1 \right]$,$G\left[ 2 \right]$,...,$G\left[ N \right]$. And when it is executed to the $j - th$ Gate $\left( {j \in \left\{ {1 \le j \le N - 1} \right\}} \right)$. According to the difference of quantum gate $G$, there are two cases:
	\begin{itemize}
		\item When $G\left[ j \right] \notin \left\{ {T,{T^\dag }} \right\}$, the server directly executes the quantum gate $G\left[ j \right]$.
		
		\item when $G\left[ j \right] \in \left\{ {T,{T^\dag }} \right\}$, where $j = {j_i}\left( {1 \le i \le M} \right)$. The server will first execute the quantum gate $G\left[ j \right]$ on the ${\omega _i}$ qubit, and then the server will perform the swap exchange operation SWAP (qubit ${\omega_i}$, qubit ${s_i}$).
	\end{itemize}
	
	\textbf{4.2. Calculate the generation of key update functions} $\left\{ {{h_i}} \right\}_{i = 1}^M$ and $f$ according to key update rules.
	
	\begin{itemize}
		\item According to key-updating rules in Fig.\ref{figure3}, the server generates the polynomial of the secret key ${a_{{j_i} - 1}}$
		\begin{equation}
		{a_{{j_i} - 1}}\left( {{\omega _i}} \right) = \left\{ {\begin{array}{*{20}{c}}
			{{h_i}\left( {{a_0},{b_0}} \right),{\kern 1pt} {\kern 1pt} \,\,\,i = 1;}\\
			{{h_i}\left( {{a_0},{b_0},{r_a}\left( 1 \right),{r_b}\left( 1 \right),...,{r_a}\left( {i - 1} \right),{r_b}\left( {i - 1} \right)} \right),\,\,i = 2,...,M}
			\end{array}} \right.
		\end{equation}
		
		\item The update function in which the server calculates the final secret key $\left( {{a_{final}},{b_{final}}} \right)$ can be expressed as:
		\begin{equation}
		f\left( {{a_{final}},{b_{final}}} \right){\rm{ = }}f\left( {{a_0},{b_0},{r_a}\left( 1 \right),{r_b}\left( 1 \right),...,{r_a}\left( M \right),{r_b}\left( M \right)} \right)
		\label{eq.18}
		\end{equation}
		
	\end{itemize}
	After the two sub-steps in step 4, the server has obtained the results of the dimensionality reduction calculation $\varphi '$, auxiliary qubits ${s_i}$, together with update-key formula $\left\{ {{h_i}} \right\}_{i = 1}^M$ and $f$. It is worth noting that the server only passes the final key calculation formula and ciphertext result to the client, and the server itself does not use any keys.
	
	\textbf{5. Measure} The client needs to alternately calculate $\left\{ {{h_i}} \right\}_{i = 1}^M$ and measure $\left\{ {{s_i},{c_i}} \right\}_{i = 1}^M$ to ensure the reliability of one secret at a time. First, the client uses the initial secret key $ek = \left( {{a_0},{b_0}} \right)$ and the measurement result ${r_a}\left( 1 \right),{r_b}\left( 1 \right)$. The specific implementation method is reviewed in Sect.\ref{2.3}
	
	The first secret key bit $a$ is calculated according to the formula $\left\{ {{h_i}} \right\}_{i = 1}^M$, and then the measurement basis $\Phi \left( {{S^a}} \right)$ is obtained. The client then measures the following qubits $\left\{ {{s_i},{c_i}} \right\}_{i = 2}^M$ to get all the measurement results  $\left\{ {{r_a}\left( i \right),{r_b}\left( i \right)} \right\}_{i = 1}^M$.
	
	\textbf{6. Decryption} $Dec\left( {ek,\left\{ {{r_a}\left( i \right),{r_b}\left( i \right)} \right\}_{i = 1}^M,f,\varphi '} \right)$. According to the initial secret key $ek = \left( {{a_0},{b_0}} \right)$ and all the measurement results $\left\{ {{r_a}\left( i \right),{r_b}\left( i \right)} \right\}_{i = 1}^M$, the final secret key $dk = \left( {{a_{final}},{b_{final}}} \right)$ is calculated by the formula $f$ in eq.\ref{eq.18}. And finally, through the decryption operation.
	\begin{equation}
	Dec\left( {dk,\varphi '} \right) = {X^{{a_{final}}}}{Z^{{b_{final}}}}\varphi ' \to \rho '
	\end{equation}
	Finally get the result we expect  $\rho '$.

\section{Experiment}
\label{5}

This section will use the IBMQ Experience to briefly describe the feasibility of this scheme. The verification will be carried out from two aspects. first, we will verify the correctness of our quantum principal component extraction algorithm on IBM Quantum experience. Due to the existence of multiple $T$-gates gates in QPCE, it is necessary to verify whether the QHE is correct when the quantum circuit is a hybrid quantum circuit of multiple $T$-gates. In addition, the IBM quantum computing platform cannot provide enough quantum registers at this stage, so we verify the dimensionality reduction scheme in cloud computing through a multi-$T$-gate hybrid quantum circuit simulation.

\subsection{Experimental implementation of Quantum principal component extraction based on IBMQ Experience}

\begin{figure*}
	\includegraphics[scale=0.25]{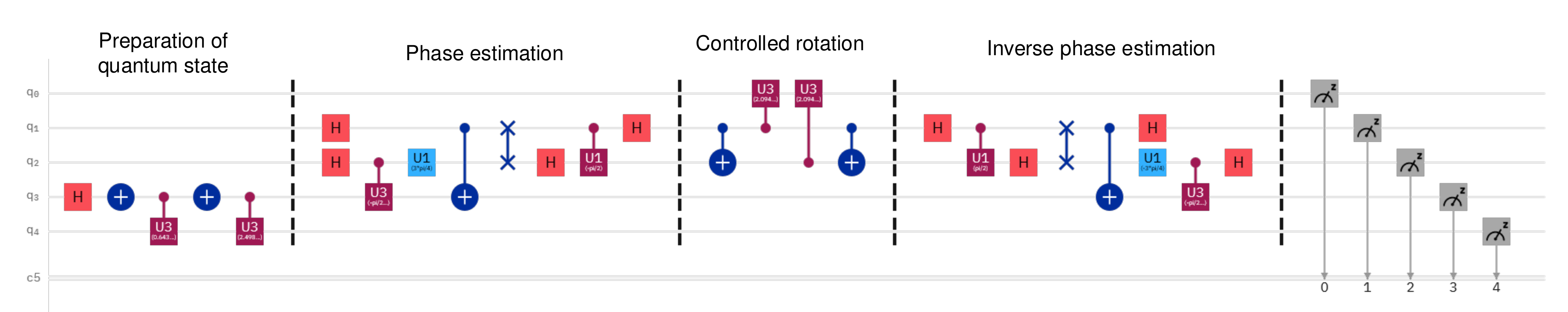}
	\caption{Quantum circuit of quantum principal component extraction. The quantum circuit includes four parts, Quantum state preparation, Phase estimation $U_{PE}$, Controlled rotation ${U_R}$ and Inverse phase estimation $U_{PE}^\dag$.}
	\label{figure7}
\end{figure*}

According to our description in Sect.\ref{3}, the quantum principal component extraction algorithm is mainly divided into four parts, Preparation of quantum state $\left| {{\varphi _\rho }} \right\rangle $, Phase estimation ${U_{{\rm{PE}}}}$, Controlled rotation ${U_R}$, Inverse phase estimation $U_{{\rm{PE}}}^\dag $. The quantum circuit is shown in the figure \ref{figure7}.

The interior of the dotted line corresponds to four parts respectively. In this experiment, it is assumed that the input density matrix $\rho {\rm{ = }}\frac{1}{2}\left[ {\begin{array}{*{20}{c}}
	3&1\\
	1&3
	\end{array}} \right]$, the threshold $\tau {\rm{ = }}0.8$. 

\textbf{Preparation of quantum state$\left| {{\varphi _\rho }} \right\rangle $ :} In the quantum circuit, we enter the normalized vector form form $\left| {{\varphi _\rho }} \right\rangle $ of $\rho $,It is normalized to get $\left| {{\varphi _\rho }} \right\rangle {\rm{ = }}{\left[ {0.670,0.223,0.223,0.670} \right]^T}$.
The quantum state is prepared by the method described in \cite{kerenidis2016quantum39}.  After calculation, the parameter of ${U_3}\left( {\theta ,0,0} \right)$ is ${\theta _1} = 0.643,\;{\theta _2} = 2.498$.

\textbf{Phase estimation ${U_{{\rm{PE}}}}$ :} This part is the same as the function implemented by HHL algorithm. The quantum circuit is designed according to \cite{2013Experimental30,2013Experimental31,2018Efficient37}. It is necessary to realize $c - {e^{ei\rho t/4}}$ and $c - {e^{ei\rho t/2}}$ in the quantum circuit and simplify it. If $t = 2\pi $ is set, it can be proved that $c - {e^{ - ei\rho t/2}} = CNOT$. At this time, ${e^{ei\rho \pi /4}} = {e^{\frac{\pi }{4}i\left( {3I + {\sigma _x}} \right)}} = {R_I}\left( { - \frac{{3\pi }}{2}} \right) \cdot {R_x}\left( { - \frac{\pi }{2}} \right)$ According to the formula in \cite{2011Quantum15}.
\begin{equation}
{R_x}\left( \theta  \right) \equiv {e^{ - i\theta X/2}} = \left[ {\begin{array}{*{20}{c}}
	{\cos \frac{\theta }{2}}&{ - i\sin \frac{\theta }{2}}\\
	{ - i\sin \frac{\theta }{2}}&{\cos \frac{\theta }{2}}
	\end{array}} \right] = \cos \frac{\theta }{2}I - i\sin \frac{\theta }{2}Y
\end{equation}

${e^{ei\rho \pi /4}}$ can be converted to

\begin{equation}
{R_x}\left( { - \frac{\pi }{2}} \right) = \cos \left( { - \frac{\pi }{4}} \right)I - i\sin \left( { - \frac{\pi }{4}} \right){\sigma _X} = \left[ {\begin{array}{*{20}{c}}
	{\frac{{\sqrt 2 }}{2}}&{\frac{{\sqrt 2 }}{2}i}\\
	{\frac{{\sqrt 2 }}{2}i}&{\frac{{\sqrt 2 }}{2}}
	\end{array}} \right] = {U_3}\left( { - \frac{\pi }{2}, - \frac{\pi }{2},\frac{\pi }{2}} \right)
\end{equation}
\begin{equation}
c - {R_1}\left( { - \frac{3}{2}\pi } \right) = {U_1}\left( {\frac{3}{4}\pi } \right)
\end{equation}
${e^{ei\rho \pi /4}}$ can be converted to ${U_3}\left( { - \frac{\pi }{2}, - \frac{\pi }{2},\frac{\pi }{2}} \right)\cdot {U_1}\left( {\frac{3}{4}\pi } \right)$. 
The rest is the detailed steps of quantum Fourier transform \cite{2011Quantum15}.

\textbf{Controlled rotation ${U_R}$ :} Because of the density matrix $\rho {\rm{ = }}\frac{1}{2}\left[ {\begin{array}{*{20}{c}}
	3&1\\
	1&3
	\end{array}} \right]$,
the eigenvalues are ${\lambda _1} = 1,\;{\lambda _2} = 2$. In addition, in order to simplify the circuit. we order $\tau {\rm{ = }}0.8$, so that $\frac{{{\lambda _1}}}{{{\lambda _2}}} = \frac{1}{2} = \left| {01} \right\rangle :\left| {10} \right\rangle $ can be converted to $\frac{{1 - \tau /{\lambda _1}}}{{1 - \tau /{\lambda _2}}} = \frac{3}{1} = \left| {01} \right\rangle :\left| {11} \right\rangle$. $\left| {10} \right\rangle  \to \left| {11} \right\rangle $ can be converted into a $CNOT$ operation on a quantum circuit. Next, the parameters for the controlled rotation of ${R_y}(\alpha )$ are set to $\alpha  = 2.094$. The method of calculation can be obtained from the optimization formula in\cite{Cong40}.

\begin{figure*}
	\includegraphics[width=0.6\linewidth]{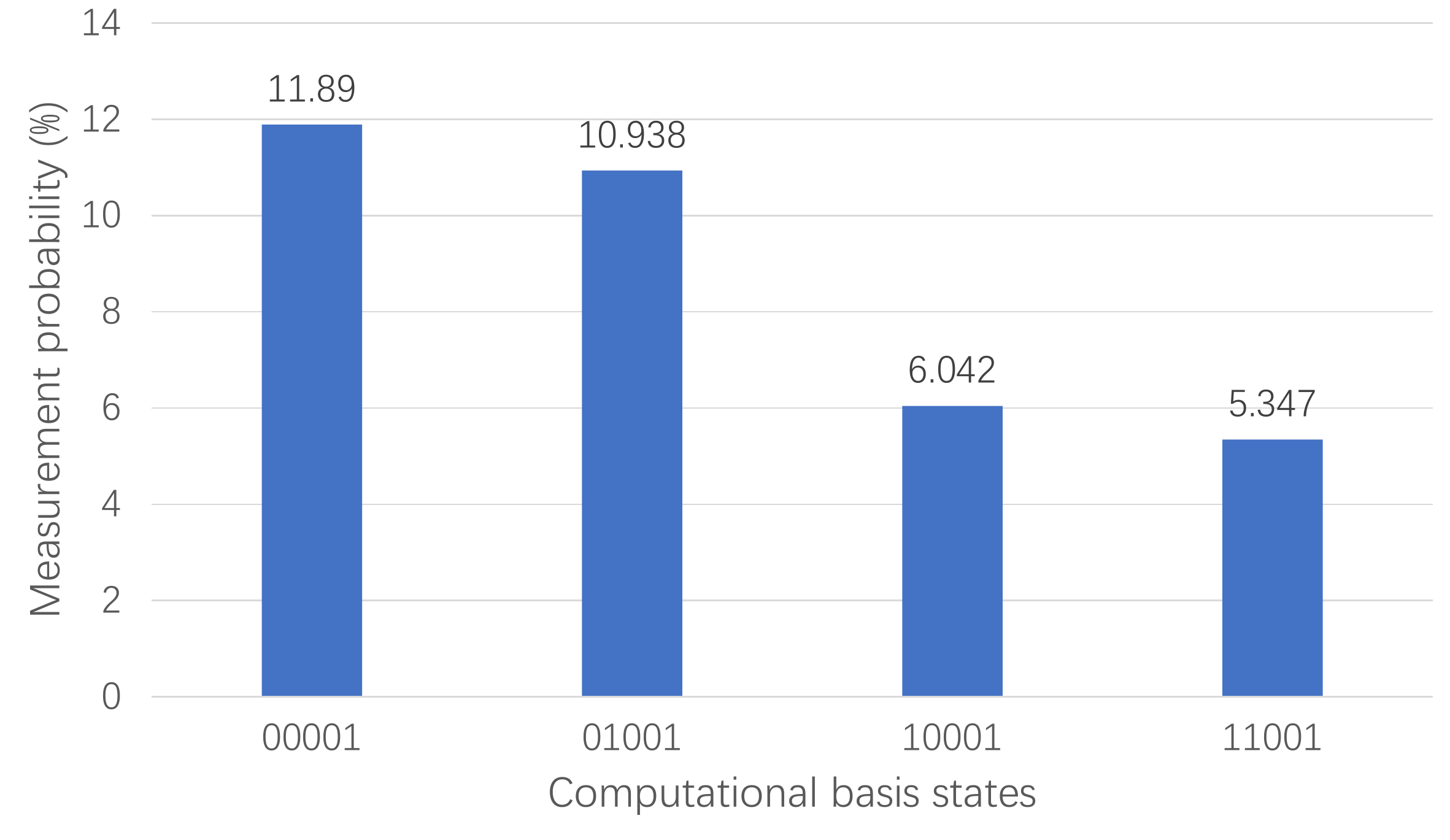}
	\caption{The results of the verification experiment of QPCE}
	\label{figure8}
\end{figure*}

The inverse phase estimation $U_{{\rm{PE}}}^\dag $ is the inverse operation of all previous operations. The measurement results of the quantum circuit are shown in the figure \ref{figure8}.

The experimental results are completed by IBM quantum computing platform, and the number of experiments is 8192. As we can see the result is
\[\begin{array}{l}
{\rho _{\rm{e}}}{\rm{ = }}{\left[ {sqrt\left( {0.1189} \right),sqrt\left( {0.10938} \right),sqrt\left( {0.06042} \right),sqrt\left( {0.05347} \right)} \right]^T}\\
\;\;\;{\kern 1pt}  = {\left[ {{\rm{0}}{\rm{.3448, 0}}{\rm{.2534, 0}}{\rm{.2312, 0}}{\rm{.3307}}} \right]^T}
\end{array}\]
According to the example, the result of the theory is ${\rho _t} = {\left[ {0.6325,0.3162,0.3162,0.6325} \right]^T}$. Calculate the normalized value of $\rho _e^ \wedge  = {\left[ {{\rm{ 0}}{\rm{.5863, 0}}{\rm{.4308, 0}}{\rm{.3932, 0}}{\rm{.5623}}} \right]^T}$. The success of the calculation results can be verified by the fidelity, that is, ${\left\| {\left\langle {\rho _e^ \wedge \left| {{\rho _t}} \right.} \right\rangle } \right\|^2} = {\rm{0}}{\rm{.9870}}$. In summary, our algorithm can be verified by experiments.

\subsection{Verify the correctness of the key update algorithm with multiple T-gates in the circuit}

\begin{figure*}
	\includegraphics[width=0.5\linewidth]{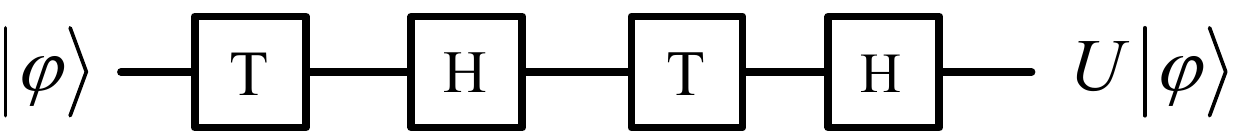}
	\caption{An example of a single-qubit quantum circuit $C$}
	\label{figure9}
\end{figure*}

\begin{figure}
	\includegraphics[width=0.8\linewidth]{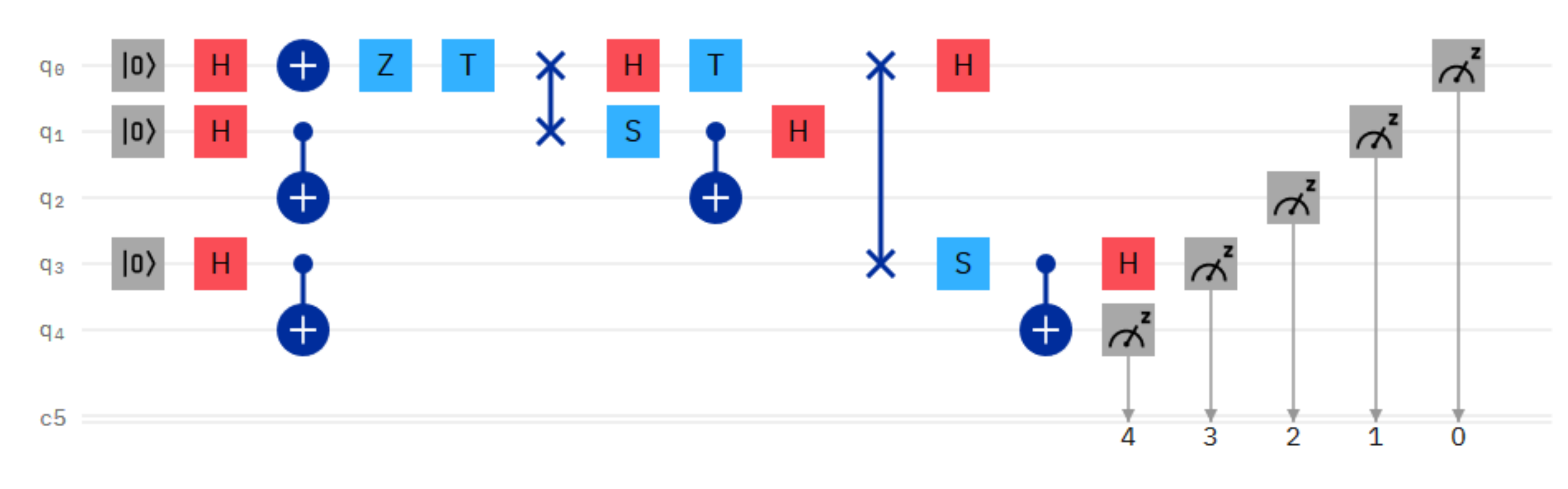}
	\caption{Verification the key update algorithm with multiple $T$-gates on IBMQ}
	\label{figure10}
\end{figure}

The verification circuit is designed according to the key update algorithm of $T$-gate and the rotated Bell basis. We design a example of quantum circuit $C$ with two $T$-gates in the figure \ref{figure9}, Where the operation of qubits $U = HTHT$. Experiments are carried out on the IBMQ cloud platform. The design quantum circuit is shown in Fig.\ref{figure10}.

According to the $QHE$ scheme, Client's initial secret key is $\left\{ {{a_0},{b_0}} \right\}$ ,The final decryption key is $\left\{ {{a_f},{b_f}} \right\}$ is calculated from the initial key and the measurement.In step 5.42,The server needs to calculate $M + 1 = 3$  key-update functions according to the key update rule.In this quantum circuit, the key- update function as follows.
\begin{equation}
{h_1}\left( {{a_0},{b_0}} \right){\rm{ = }}\;{a_0}
\end{equation}
\begin{equation}
{h_2}\left( {{a_0},{b_0},{r_a}\left( 1 \right),{r_b}\left( 1 \right)} \right){\rm{ = }}\;{a_0} \oplus {b_0} \oplus {r_b}\left( 1 \right)
\end{equation}

The final Key-updating function is
\begin{equation}
\left\{ {{a_f},{b_f}} \right\} = f\left( {{a_0},{b_0},{r_a}\left( 1 \right),{r_b}\left( 1 \right),{r_a}\left( 2 \right),{r_b}\left( 2 \right)} \right)
\end{equation}
where
${a_f} = \;{b_0} \oplus {r_a}\left( 1 \right) \oplus{r_b}\left( 1 \right)\oplus {r_b}\left( 2 \right)$,
${b_f} = \;{a_0} \oplus {b_0} \oplus {r_b}\left( 1 \right) \oplus {r_a}\left( 2 \right)$.

In the decryption process of the quantum circuit, the client needs to carry out two steps of operation.
\begin{itemize}
	\item When the key update function ${h_1}{\rm{ = }}1$, the client executes the measurement basis on the first T-gates $\Phi \left( {{S^{{a_0}}}} \right)$, and measure it to get a pair of qubits $\left( {{r_a}\left( 1 \right),{r_b}\left( 1 \right)} \right)$.
	\item When the key update function ${h_2}{\rm{ = }}1$, the client executes the measurement basis on the first T-gates $\Phi \left( {{S^{{a_f} = \;{b_0} \oplus {r_a}\left( 1 \right) \oplus{r_b}\left( 1 \right)\oplus {r_b}\left( 2 \right)}}} \right)$ and measure it to get a pair of qubits $\left( {{r_a}\left( 2 \right),{r_b}\left( 2 \right)} \right)$.
\end{itemize}

Finally, according to the ${r_a}\left( 1 \right),{r_b}\left( 1 \right),{r_a}\left( 2 \right),{r_b}\left( 2 \right)$ measured by the client above and the key-update function $f$ calculated by the server, the client gets the final key $\left\{ {{a_f},{b_f}} \right\}$ and decrypts the result to get the final result.

\begin{figure}
	\includegraphics[width=0.92\linewidth]{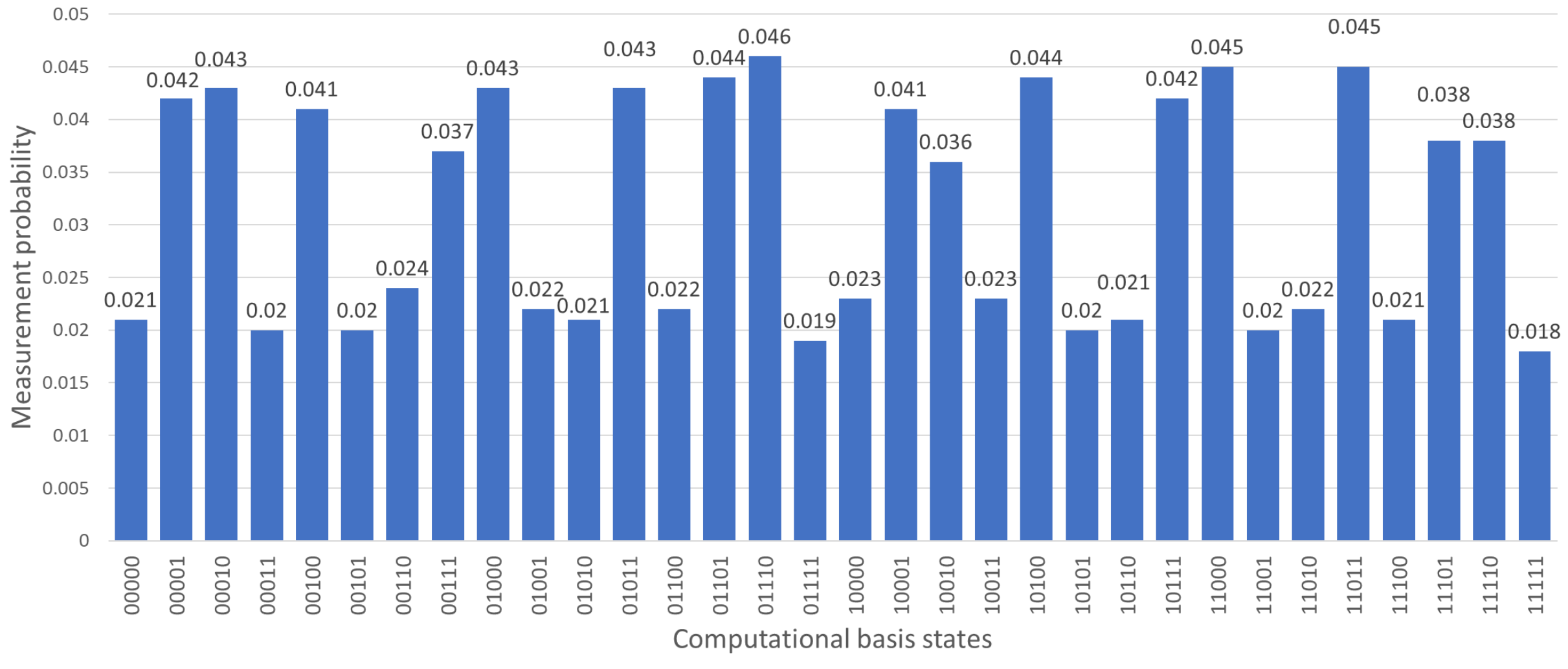}
	\caption{The measurement result of verifying the quantum homomorphic encryption scheme is based on IBMQ}
	\label{figure11}
\end{figure}

This experiment is carried out by using IBMQ computing platform located in melbourne (ibmq\_16\_melbourne). In order to ensure the accuracy of the experiment, the highest number of runs is 8192. The plaintext ${q_0}$ in the circuit is set to $\left| 0 \right\rangle $ . Set encryption initial key $EK = \left\{ {{a_0},{b_0}} \right\}{\rm{ = }}\left\{ {1,1} \right\}$. $\left\{ {q1,q2} \right\},\left\{ {q3,q4} \right\}$ is two sets of auxiliary Bell state. According to the definition of rotated Bell basis, we perform ${S^a}$ rotation measurement on $\left\{ {q1,q2} \right\},\left\{ {q3,q4} \right\}$. The measurement results are ${r_a(2),r_b(2),r_a(1)},{r_b(1)}, {q_0}$ from left to right. The running results of the experiment are as follows in figure \ref{figure11}. 

Obviously, the result we want will be displayed with high probability. Four results of $\left\lbrace  00001,00010,\right.\\\left.01011,01110\right\rbrace$ are randomly selected as examples in the measurement results, and the key-update process is shown in Table \ref{table1}.

\begin{table}
	\caption{Updated keys and measurement result}
	\label{table1}
	\scalebox{0.85}{
		\begin{tabular}{ccccccc}
			\toprule
			${r_a}\left( 2 \right)$ & ${r_b}\left( 2 \right)$ & ${r_a}\left( 1 \right)$ & ${r_b}\left( 1 \right)$& ${b_0} \oplus {r_a}\left( 1 \right) \oplus{r_b}\left( 1 \right)\oplus {r_b}\left( 2 \right)$ & ${a_0} \oplus {b_0} \oplus {r_b}\left( 1 \right) \oplus {r_a}\left( 2 \right)$ & measurement result of ${q_0}$\\
			\midrule
			0 & 0 & 0 & 0 & 1 & 0 & 1\\
			0 & 0 & 0 & 1 & 0 & 1 & 0\\
			0 & 1 & 0 & 1 & 1 & 1 & 1\\
			0 & 1 & 1 & 1 & 0 & 1 & 0\\
			\bottomrule
		\end{tabular}
	}
\end{table}
First, we execute the quantum circuit shown in Fig. \ref{figure9} in plaintext, and the result of ${q_0}$ is 0. We use the updated key to decrypt the ciphertext, according to the decryption formula.

\begin{equation}
{X^{{a_{f}}}}{Z^{{b_{f}}}}\varphi ' \to \rho '
\end{equation}
We use the decrypted final key $ \left\lbrace  {a_f},{b_f}\right\rbrace $ to decrypt the result of $q_0$ measurement. We can get a phenomenon. If ${a_f} = \;{b_0} \oplus {r_a}\left( 1 \right) \oplus{r_b}\left( 1 \right)\oplus {r_b}\left( 2 \right)=1$, the measurement result of ${q_0}$ is always 1. On the other hand, If ${a_f} = 0 $,  the measurement result of ${q_0}$ is always 0. In this way, we can always decrypt it and get ${q_0}=0$. In summary, our scheme can be proved to be correct.

Compared with the verification test of a single t-gates verified in\cite{2020Grover38}, we can conclude that when there are more $T$-gates in the circuit, as long as we strictly follow the quantum key-update algorithm, we can get the correct decryption key.

\section{Performance analysis}
\label{6}
The quantum principal component extraction algorithm implemented in this paper completely retains the main framework of HHL-like algorithm. The completeness and correctness of the algorithm are realized by quantum subroutines: quantum phase estimation, controlled rotation and inverse quantum phase estimation. While ensuring the correctness of quantum principal component extraction, it is also necessary to ensure the correctness, security and quasi-compactness of the encryption process.

1.This scheme has been proved to be correct\cite{2019QuIP...19...28L35}. The scheme points out that if the logic gates in the quantum circuit in the evaluation function in the encryption process are all composed of 
\begin{equation}
S = \left\{ {X,Z,H,S,P,CNOT,T,{T^\dag }} \right\}
\end{equation}
Then the GT scheme is a fully homomorphic encryption scheme. In this scheme, the algorithm not only has these quantum logic gates, but also has revolving gates ${R_y}$. But in \cite{2011Quantum15} the formula (4.4)
\begin{equation}
{R_y}\left( \theta  \right) \equiv {e^{ - i\theta Y/2}} = \left[ {\begin{array}{*{20}{c}}
	{\cos \frac{\theta }{2}}&{ - \sin \frac{\theta }{2}}\\
	{\sin \frac{\theta }{2}}&{\cos \frac{\theta }{2}}
	\end{array}} \right] = \cos \frac{\theta }{2}I - i\sin \frac{\theta }{2}Y
\end{equation}

where $\theta $ is the parameter. It can be seen that it is simply derived from the unit matrix $I$ and the pauli Y matrix. Therefore, it also meets the correctness requirements of the GT scheme\cite{liang2020teleportation(21)}. For arbitrary n-qubit quantum circuits $C$ and n-qubit qubit data $\rho $. There is a decryption process $Dec\left( {ek,\left\{ {{r_a}\left( i \right),{r_b}\left( i \right)} \right\}_{i = 1}^M,f,Eval\left( {C,\left\{ {{s_i}} \right\}_{i = 1}^M,\varphi' } \right)} \right)$, where $C$ is composed of General quantum gate $G\left[ 1 \right]$, $G\left[ 2 \right]$,..., $G\left[ N \right]$ and $M = \left\{ {j|G\left[ {{j_i}} \right] \in \left\{ {T,{T^\dag }} \right\}} \right\}$. 

In this scheme, we use the rotary measurement operation $\Phi \left( {{S^{{a_{{j_i} - 1}}\left( {{\omega _i}} \right)}}} \right)$ to eliminate the errors introduced by the T gate in the encryption process, so that it can correctly evaluate the homomorphism, and   represents the currently encrypted quantum T gate $G\left[ {{j_i}} \right]$,.

The rotary measuring operation $\Phi \left( {{S^{{a_{{j_i} - 1}}\left( {{\omega _i}} \right)}}} \right)$ contains two operations completed in $Eval$ and $Dec$: SWAP——the exchange operation and the measurement operation. The measurement of $Dec$ mainly depends on whether an intermediate qubit is correct or not. If it is correct, it can be executed normally.

According to the fourth step in the scheme, we can know that because the order of the quantum gates is fixed, we can postpone the measurement of $\Phi \left( {{S^{{a_{{j_i} - 1}}\left( {{\omega _i}} \right)}}} \right)$ and have no effect on the final result. Therefore, it is assumed that the operational measurements of both parts of $\Phi \left( {{S^{{a_{{j_i} - 1}}\left( {{\omega _i}} \right)}}} \right)$ are carried out accurately in $Eval$ and accurate results are obtained $\left\{ {{r_a}\left( 1 \right),{r_b}\left( 1 \right)} \right\},...,\left\{ {{r_a}\left( M \right),{r_b}\left( M \right)} \right\}$. 

In this way, in the process of $Eval$, when we implement our quantum principal component extraction for plaintext, we encrypt it in turn according to the quantum gate order we designed in advance. And ensure that every time the quantum gate is executed, the quantum secret key update algorithm is strictly implemented, and the intermediate secret key-updating function $\left\{ {{h_i}} \right\}_{i = 1}^M$ and the final key-updating function $f$ are obtained. Finally, in the $Dec$, the intermediate key ${\left( {{a_i},{b_i}} \right)_M}$ and the final key 
$\left( {{a_{final}},{b_{final}}} \right)$ calculated in $Eval$ are correct. After the decryption step: 
\begin{equation}
Dec\left( {dk,\varphi '} \right) = {Z^{{a_{final}}}}{X^{{b_{final}}}}\varphi '{X^{{b_{final}}}}{Z^{{a_{final}}}}
\end{equation}

The final decryption gets the correct result $\rho '$

2.This scheme is perfectly safe. In the encryption process, the encryption scheme can perfectly hide our initial key $ek = \left( {{a_0},{b_0}} \right)$ and plaintext $\rho $. As long as you ensure that keys   and   randomly select a set of keys consisting of 0,1. And the client uses the combination of quantum gate pauli X and pauli Z to act on plaintext $\rho \;\left( {\rho  \to \varphi {\kern 1pt} \,{\rm{ = }}\,{X^a}{Z^b}\rho } \right)$. For the input plaintext quantum state $\rho $ and the output ciphertext quantum state $\rho '$ is the maximum mixed state.
\begin{equation}
\frac{1}{{{2^{2n}}}}\sum\limits_{a,b \in {{\left\{ {0,1} \right\}}^n}} {{X^a}{Z^b}\rho {{\left( {{X^a}{Z^b}} \right)}^\dag }}  = \frac{{{I_{{2^n}}}}}{{{2^n}}}
\end{equation}

Where $\frac{{{I_{{2^n}}}}}{{{2^n}}}$ is the maximum mixed state. According to this formula, Boykin and Roychowdhury have also been proved in quantum one-time pad(QOTP) \cite{2003Optimal36}. And there is no interaction between the client and server during the evaluation process. its benefits from the quantum one-time pad(QOTP) encryption transformation we use also. Makes it impossible for the server to know any information about the key and plaintext throughout the execution of the scheme. In the process of performing decryption, the client only carries out some classical calculations and multiple quantum measurements, and does not interact with the client, and the measurement itself is a part of the decryption. Therefore, this scheme is perfectly safe.

3.The scheme itself is quasi-compact. Since every key update formula can be regarded as a binary number composed of XOR gates, according to our update rule, we can see:
\begin{equation}
\left\{ \begin{array}{l}
{h_i}:{\left\{ {0,1} \right\}^{2n + 2(i - 1)}}{\kern 1pt}  \to \left\{ {0,1} \right\},i = 1,...,M;\\
f:{\left\{ {0,1} \right\}^{2n + 2M}}{\kern 1pt}  \to {\left\{ {0,1} \right\}^{2n}}.
\end{array} \right.
\end{equation}

The maximum computational complexity of the decryption process can be obtained as follows:
\begin{equation}
\sum\nolimits_{i = 1}^M {{{\log }_2}\left( {2n + 2(i - 1)} \right)}  + 2n{\log _2}\left( {2n + 2M} \right) = O\left( {\left( {M + n} \right){{\log }_2}\left( {M + n} \right)} \right)
\end{equation}

Where the computational complexity of ${h_i}$ is ${\log _2}2n + 2(i - 1)$. The measurement operation in the decryption process requires M measurements and n-bit decryption, and the quantum complexity of the operation is $M + 2n$. Therefore, the decryption process has the quasi-compactness of $M\log M$ to the quantum circuit.

\section{Conclusion and Future work}
\label{7}
In this paper, a new quantum principal component extraction (QPCE) is proposed. Compared with the classical algorithm, this algorithm has an exponential improvement. Combined with the quantum homomorphic encryption scheme, a quantum homomorphic ciphertext dimension reduction scheme (QHEDR) is proposed for the first time. When the amount of data is large, plaintext can be calculated on the cloud (server) by once encryption, and there is no need for key-interaction. The client only requires $M$ measurements and QOTP key update of $n$-bit quantum to get the calculated results. The scheme ensures security, correctness and quasi-compactness.

In addition, an example is given to verify the circuit of the quantum principal component extraction algorithm on the IBMQ computing platform. However, the complexity of the circuit $T$-gate of QPCE is high, so the decryption process of this scheme is more complex. And when there is a $T$-gate, two auxiliary quantum circuits are needed. The quantum circuits supported by the IBMQ computing cloud platform are insufficient. Therefore, we verify the correctness of the hybrid quantum circuit to replace the correctness of the complete QPCE under the ciphertext. The experimental results show that when there are $T$-gates in the quantum circuit and there are enough auxiliary quantum circuit, it is feasible to implement QHEDR scheme.

The proposal of the quantum computing scheme provides a solution and idea for the privacy problem in the cloud. In the future, we will implement this scheme on a real quantum circuit to reduce the complexity of $T$-gate and optimize the encryption scheme, so as to reduce the workload of the client and server.

\begin{acks} 
This research was funded by the Scientific Research Foundation for Advanced Talents from Shenyang Aerospace University -18YB06, and National Basic Research Program of China - JCKY2018410C004.	
\end{acks}

\bibliographystyle{ACM-Reference-Format}
\bibliography{sample-base}

\end{document}